\documentclass[prb,twocolumn,floatfix]{revtex4}

\usepackage[T1]{fontenc}
\usepackage[latin1]{inputenc}
\usepackage{amsmath,amssymb,amsfonts,graphicx}

\renewcommand{\dag}{^{\dagger}}
\newcommand{\ket}[1]{\lvert #1 \rangle}
\newcommand{\bra}[1]{\langle #1 \rvert}
\newcommand{\exv}[1]{\langle#1\rangle}
\newcommand{\eps}{\epsilon}
\newcommand{\sig}{\sigma}
\newcommand{\hf}{^{\text{HF}}}

\renewcommand{\a}{\alpha}
\renewcommand{\b}{\beta}
\newcommand{\m}{\mu}
\newcommand{\n}{\nu}
\renewcommand{\dag}{^{\dagger}}

\renewcommand{\Im}{\mathrm{Im}}

\begin{document}

\title{Charge and spin response functions for the Tomonaga model with quadratic dispersion and different interactions}
\author{Patrick Pl\"otz\footnote{p.ploetz@thphys.uni-heidelberg.de}}
\affiliation{\mbox{Institut f\"ur Theoretische Physik, Universit\"at Heidelberg, Philosophenweg 19, 69120 Heidelberg, Germany}}

\begin{abstract}
We derive expressions for the charge and spin response function for the Tomonaga model with quadratic dispersion and arbitrary (but finite for zero momentum) interaction. For constant interaction these expressions are analytic and for other types of interaction only a simple matrix has to be diagonalised. We use a truncated expansion in particle-hole states with and without inclusion of correlations in the ground state yielding an exact result for pure intra-band interaction. We also discuss the possibility of power-laws in the dynamic structure factor for the spinful and spinless model.
\end{abstract}

\maketitle
\section{Introduction}
Low dimensional systems of interacting electrons are an active field of research, with the interaction playing a more dominant role than in higher dimensions. In his famous 1950 work Tomonaga~\cite{Tomonaga} found the exact solution of such a model system for one spatial dimension based on the following assumption: If the range of the interaction is much larger than the inter-particle distance, the quadratic energy dispersion can be \emph{linearised} around the Fermi points. Tomo\-naga found collective excitations of the electronic density, i.e. plasmons as the low-lying excitations of such a system. The systematic extension of this model is nowadays called Tomonaga-Luttinger (TL) model and discussed in many reviews\cite{Voit} and books.\cite{Giamarchi}

Quite recently, several authors attempted to go beyond this approximation and to include effects of the non-linearity in the energy dispersion.\cite{Kopietz, Russen, Teber, Spins, KS} Of main interest was the shape of the dynamic structure factor (DSF) where the plasmons of the TL model show up as simple delta-peaks and changes are expected as broadening of the peak\cite{Kopietz} (corresponding to plasmon damping) or as the appearance of power-laws.\cite{Spins, Russen, Teber}\\
A possible starting point for a discussion could be given by the \emph{random phase approximation}~(RPA) which is nothing else but linearised time-dependent Hartree theory with the Hartree-Fock energies replaced by the non-interacting ones. This approximation becomes exact in the limit of linear dispersion, yielding the exact solution of the TL model. But it has been shown\cite{KS, Diplomarbeit} only recently as an insufficient starting point for a discussion of the above mentioned questions. An analytic solution for the special case of a constant interaction has been derived within a better approximation -- the random phase approximation with exchange (RPAE) corresponding to linearised time-dependent Hartree-Fock theory (TDHF). 

The linearization of the dispersion leads to two linear branches instead of a single quadratic one. Electrons on the right branch are usually called right-movers and those on the left branch left-movers. Accordingly, one can distinguish interactions between electrons on the same and on different branches. In this work we make explicit use of this distinction between inter-band and intra-band interactions. The shape of the interaction under consideration is arbitrary with only one constraint necessary for Luttinger liquid phenomenology: the Fourier components of the interaction are finite and only non-zero for small momenta.\cite{Giamarchi} After introducing response functions in general in Sec.~\ref{sec:RF}, we focus on pure intra-band interactions in Sec.~\ref{sec:g4}. The ground state turns out to be uncorrelated (a Slater determinant) and excited states can be expanded in particle-hole states. We derive simple expressions for the response functions in the spinful and spinless case which includes the exact solution for small momenta and is analytic for constant interaction. In the following Sec.~\ref{sec:full_model} also inter-band interactions are taken into account on the RPAE level. For both models the transition in the DSF between a curvature dominated limit akin to the non-interacting case and the limit of strong interaction close to the TL model are discussed. We also consider the possibility of power-law divergences in an intermediate regime. We close this article with concluding remarks in Sec.~\ref{sec:summary}. 

\section{Response functions}\label{sec:RF}
We consider interacting electrons on a ring of length $L$. In units where $\hbar=1$ this is described by a Hamiltonian 
	\begin{equation}H=\sum_k\epsilon_kc_k\dag c_k^{} + \frac{1}{2}\sum_{k_1k_2k_3k_4}v_{k_1,k_2;k_3,k_4}c_{k_1}\dag c_{k_2}\dag c_{k_4}^{}c_{k_3}^{},\end{equation}
where $\eps_k=k^2/2m$ denotes the non-relativistic quadratic dispersion and $v_{k_1,k_2;k_3,k_4}$ the matrix elements of the two-body interaction, containing a factor $\delta_{k_1+k_2,k_3+k_4}$ ensuring momentum conversation. A possible spin index $\sig$ is included in $k$.

The quantities of interest are, firstly, the (generally spin-dependent) \emph{density-density response function}\cite{Giuliani}
\begin{equation}\chi^{\sig\sig'}(x-x',t)\equiv-i\theta(t)\exv{[\rho_{\sig}(x,t),\rho_{\sig'}(x',0)]},\end{equation}
where $\rho_{\sig}(x)$ is the density operator and $\sig=\pm$, which leads to a \emph{charge} and \emph{spin response function} in the spinful model denoted by 
\begin{gather}\chi^{(c)}_{}\equiv \sum_{\sig\sig'}\chi^{\sig\sig'}\text{ and }\chi^{(s)}_{} \equiv \sum_{\sig\sig'}\sig\sig' \chi^{\sig\sig'}, \end{gather} 
respectively. Secondly, the \emph{dynamic structure factor} $S(q,\omega )$ being related to the response function by 
$-\pi^{-1}\Im \chi(q,\omega )=S(q,\omega )$ at $T=0$ and by the fluctuation dissipation theorem in general.\cite{Landau} 
Switching to Fourier space and using a spectral decomposition we introduce spin-dependent particle-hole (\emph{ph}) states 
\begin{equation}
	\ket{q,\sig}\equiv \rho_{-q}^{\sig}\ket{0}=\sum_kc\dag_{k+q,\sig}c^{}_{k,\sig}\ket{0}, 
\end{equation}
where $\ket{0}$ denotes the ground state to $H$, and get for the response function at $T=0$
\begin{equation}\label{eq:Chi}
\chi^{\sig\sig'}(q,\omega ) =\sum_n\left[\frac{\bra{q,\sig}n\rangle\bra{n}q,\sig'\rangle}{\omega -[E_n-E_0]}-\frac{\bra{q,\sig'}n\rangle\bra{n}q,\sig\rangle}{\omega +[E_n-E_0]}\right]
\end{equation}
where $\omega  = \omega +i0$ and $\ket{n}$ denotes a complete set of eigenstates of $H$. 
Introducing states $\ket{s}=\ket{+}+\ket{-}$ and $\ket{a}=\ket{+}-\ket{-}$ the charge and spin response functions can be written as \begin{equation}\chi^{(c)}_{}(q,\omega )=\bra{s}\chi^{\sig\sig'}\ket{s}\text{ and }\chi^{(s)}_{}(q,\omega )=\bra{a}\chi^{\sig\sig'}\ket{a}. \end{equation} 
These quantities shall subsequently be calculated in different approximations.

\section{Pure intra-band interactions}\label{sec:g4}
\subsection{$g_4$-model and Tamm-Dancoff approximation}
Following Tomonaga, we assume the range of the interaction to be much larger than the inter-particle distance or, equivalently, the Fourier components of the interaction are non-zero only for small momenta (smaller than some cut-off $k_c$), $v(k)\neq0$ for $k\leq k_c\ll k_F$. The interaction leads to excitations around the Fermi points and one distinguishes two types of electrons: right(left)-movers corresponding to electrons with momenta close to $+(-)k_F$. We introduce g-ology notation\cite{solyom} and denote interactions involving only one sort of electrons as $g_4$-processes (intra-band interactions) and those between different types of electrons as $g_2-$processes (inter-band interactions). Using this distinction the Hamiltonian is a sum of the kinetic energy $H_0$ and interactions $H_2$ and $H_4$. Interestingly, for pure intra-band interactions the ground state of the interacting system is still the Fermi sea $\ket{\text{FS}}$, as can be seen by direct application of $H_4$ to $\ket{\text{FS}}$. 

We will now focus on pure $g_4$-interaction and drop this limitation in Sec.~\ref{sec:full_model}. Within this frame we have complete knowledge of the ground state and the exact excited states can systematically be expanded in particle-hole states. Such a truncated expansion up to 1-\emph{ph} contributions is widely used in nuclear physics and called Tamm-Dancoff approximation~(TDA).\cite{RingSchuck} We write the ground state as a Hartree-Fock (HF) state $\ket{0}=\ket{\text{HF}}$, which coincides with $\ket{\text{FS}}$ for the $g_4$-model, and consider excited states containing 1- and 2-\emph{ph}--contributions
\begin{equation}\label{eq:TDA}
	 \ket{n}=\ket{\text{HF}}+\sum_{\alpha\mu}c\dag_{\alpha}c^{}_{\mu}\ket{\text{HF}} +\sum_{\alpha\beta\mu\nu}c\dag_{\alpha}c\dag_{\beta}c^{}_{\mu}c^{}_{\nu}\ket{\text{HF}}
\end{equation}
where $\alpha, \beta, \ldots$ denote \emph{unoccupied} ($|k|>k_F$) and $\mu, \nu, \ldots$ \emph{occupied} ($|k|\leq k_F$) one-particle states. For not too strong interactions the 1-\emph{ph} excitations are most important and higher orders yield only small corrections. 
Another important point for estimating the approximation is the following. Using periodic boundary conditions the momentum is quantised, $q=m_q\cdot L/(2\pi)$. Thus, e.g. for 2-\emph{ph} excitations, only those combinations of momenta for which $\a+\b-\mu-\nu = q$ are involved. The overall number of possible $n$-\emph{ph} excitations is the number of possible partitions of a given number $m_q$ into a sum over $n$ integer particle momenta, which is usually referred to as partition function in the mathematical literature.\cite{AS} 
For small $m_q$ the single and double \emph{ph} excitations span the most important part of the Hilbert space. In particular for $m_q=8$ only 1- and 2-\emph{ph}-excitations are possible and our approximation Eq.~(\ref{eq:TDA}) yields the \emph{exact} result for pure intra-band interactions. For $m_q=20$, which is the value for most of the numerical results in this paper, these excitations make up half of the states spanning the full Hilbert space and higher excitations give only small corrections.

\subsection{Results}
\begin{figure}[t]
	\centering
	\includegraphics[width=0.95\linewidth, height=0.25\textheight]{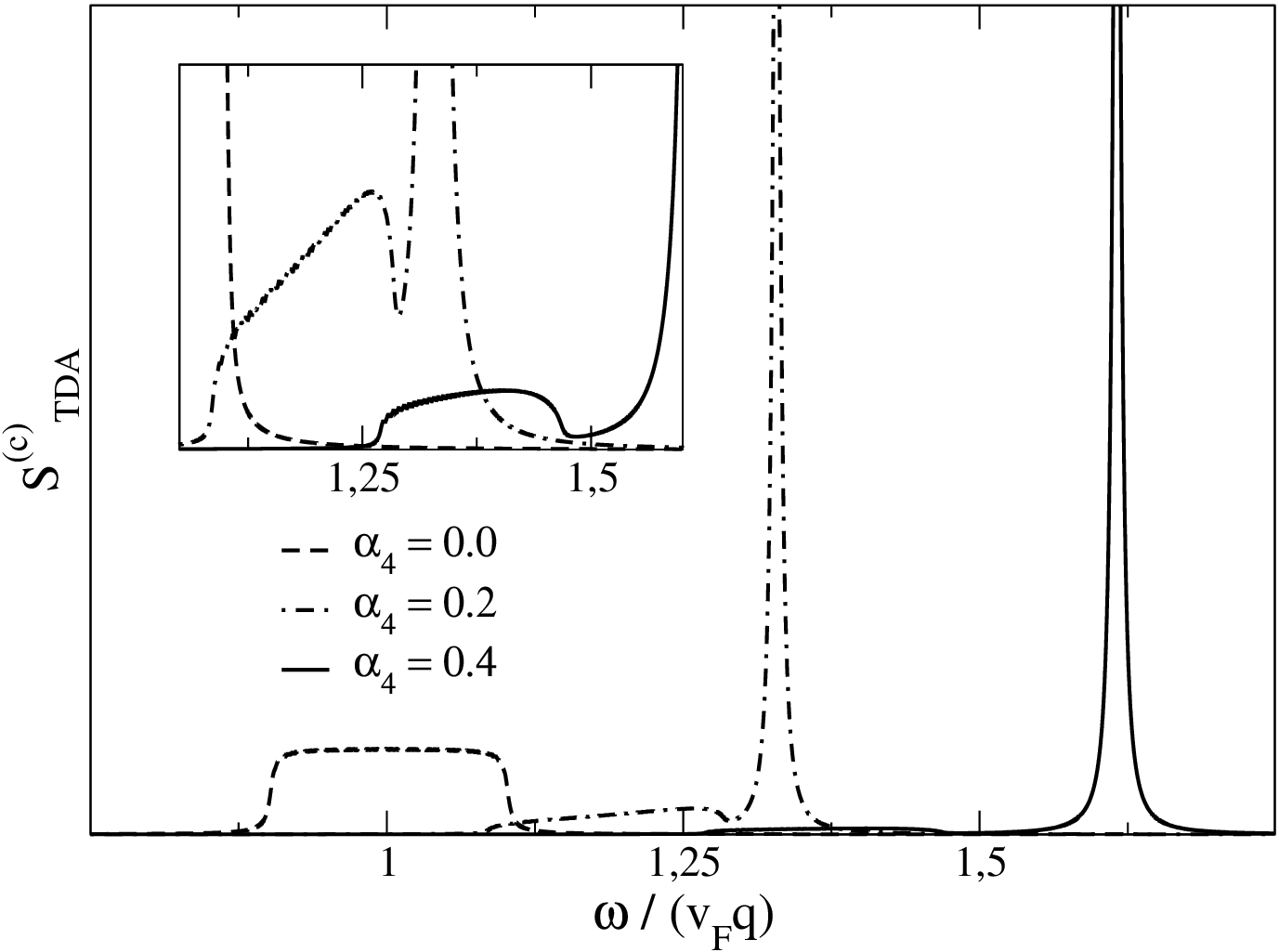}
	\caption{Charge response function in TDA for $\a_4=0.0$ (dashed), $\a_4=0.2$ (dashed-dotted) and $\a_4=0.4$ (full line) for quadratic $g_4(q)=g_4(0)[1-q^2/k_c^2]$ interaction and constant momentum $\tilde q= q/(2k_F)=0.1$. The HF-continuum moves to higher energies with growing interaction strength and looses spectral weight compared to the emerging peak (inset). Calculations for $m_q=2\pi q/L=50$. The finite width of the peaks results from a finite imaginary part $\eta = 2.5\cdot10^{-3}$ which leads to a sum of Lorentzians instead of delta peaks as would be the case for $\eta=0$.}
	\label{fig:LR_TDA}
\end{figure}

In a first step we consider single \emph{ph} excitations and extend this approximation in the next paragraph.
To make use of Eq.~(\ref{eq:Chi}) we have to calculate \emph{ph}-matrix elements of $H-E_0$, where $E_0=E_0\hf$ since the ground state is a Slater determinant as explained above. Using $\bra{\text{HF}}c_{\mu'}\dag c_{\alpha'}^{}(H-E_0\hf)c_{\alpha}\dag c_{\mu}\ket{\text{HF}} = \bra{\text{HF}}c_{\mu'}\dag c_{\alpha'}^{}[H,c_{\alpha}\dag c_{\mu}]\ket{\text{HF}}$, one finds 
\begin{equation*}\bra{\alpha'\mu'}(H-E_0\hf)\ket{\alpha\mu}=(\eps_{\alpha}\hf-\eps_{\mu}\hf)\delta_{\alpha\alpha'}\delta_{\mu\mu'} - \bar v_{\alpha',\mu;\alpha,\mu'},\end{equation*} 
where we introduced $\ket{\alpha\mu}\equiv c\dag_{\alpha}c^{}_{\mu}\ket{\text{HF}}$ as a short hand notation and HF energies $\eps_k\hf \equiv \eps_k+\sum_{k'}\bar v_{k,k';k,k'}f(\eps_{k'})$. At $T=0$ the Fermi function $f(\eps)$ is a simple step function $\Theta(k_F-|k|)$. Writing momenta and spin indices explicitly, this reads 
	\begin{multline}\bra{\text{HF}}c_{k',\sig'}\dag c_{k'+q,\sig'}^{}(H-E_0\hf)c\dag_{k+q,\sig}c^{}_{k,\sig}\ket{\text{HF}} =\\
	(\eps_{k+q}\hf-\eps_k\hf)\delta_{kk'}\delta_{\sig\sig'} +\frac{1}{L}[g_4(q)-\delta_{\sig\sig'}g_4(k'-k)]
	\end{multline}
where the Fourier components $g_4(k)$ of the interaction are assumed to be non-vanishing only for small momenta $|k|\leq k_c\ll k_F$.
The basic relation Eq.~(\ref{eq:Chi}) can now be written as an equation for the matrix elements of the response function
	\begin{equation}\begin{split}
	 \sum_{l\tilde\sig}\Bigl\{[\omega-(\eps_{k+q}\hf-\eps_k\hf)]\delta_{kl}\delta_{\sig\tilde\sig} \hspace{2cm}\\
	-[g_4(q)-\delta_{\sig\tilde\sig}g_4(k-l)]\Bigr\} \chi_{lk'}^{\tilde\sig\sig'} = \delta_{kk'}\delta_{\sig\sig'}.\end{split}
	\end{equation}
or equivalently in form of a matrix equation
	\begin{equation}\label{eq:Matrix_g4}
	\left(\begin{array}{cc}\omega-a & -b_4\\-b_4 & \omega -a \end{array}\right)
	\left(\begin{array}{cc}\chi^{++} & \chi^{+-}\\\chi^{-+} & \chi^{--} \end{array}\right)
	= \left(\begin{array}{cc} 1 & 0 \\ 0 & 1 \end{array}\right)
	\end{equation}
where $(a)_{k,k'}=(\eps_{k+q}\hf-\eps_k\hf)\delta_{kk'} + [g_4(q)-g_4(k-k'')]$ and $(b_4)_{k,k'}=g_4(q)$. 
The charge and spin response function are therefore given by
	\begin{equation}\begin{split}\label{eq:ChiTDA}
	\chi^{(c)}_{g_4,\pm} (q,\omega)  = \sum_{\sig\sig'}\chi^{\sig\sig'} = \frac{2\chi_{\pm}}{1-g_4(q)\chi_{\pm}} \\
	\chi^{(s)}_{g_4,\pm} (q,\omega)  = \sum_{\sig\sig'}\sig\sig'\chi^{\sig\sig'} = \frac{2\chi_{\pm}}{1+g_4(q)\chi_{\pm}},\end{split}
	\end{equation}
leading to $\chi_{\text{TDA}}^{(c/s)}=\chi^{(c/s)}_{g_4,+}+\chi^{(c/s)}_{g_4,-}$ with the HF response functions $\chi_{\pm}(q,\omega)\equiv\mp\sum_{kk'}[(\omega\pm a)^{-1}]_{kk'}$. For a constant interaction $g_4(q)=g_4(0)\Theta(k_c-|q|)$ the latter are given by 
	\begin{equation}
	\chi_{\pm}(q,\omega) = -\frac{1}{2\tilde q}\frac{1}{2\pi v_F}\ln\bigg|\frac{1+\alpha_4+\tilde q \mp \tilde\omega}{1+\alpha_4-\tilde q \mp\tilde\omega}\bigg|
	\end{equation}
but in general have to be determined by diagonalisation of $a$. The dimensionless quantities 
	\begin{equation*}
	\alpha_i=\frac{g_i(0)}{2\pi v_F}, \quad \tilde q = \frac{q}{2k_F}, \quad \tilde\omega = \frac{\omega}{v_Fq}
	\end{equation*}
will be used throughout this paper. 
Figure~\ref{fig:LR_TDA} shows the charge response function in 1-\emph{ph}-TDA for fixed momentum $q/(2k_F)=0.1$ and increasing interaction strength. A transition from the box-like shape in the non-interacting case to a single delta-peak as known from the TL model is apparent. The HF-continuum moves to higher energies with growing interaction and looses spectral weight to the plasmon peak. For constant interaction it is non-vanishing for frequencies $1+\a_4-\tilde q\leq \tilde\omega \leq 1+\a_4 +\tilde q$. 
\\
For the spinless case the block matrix equation~(\ref{eq:Matrix_g4}) collapses to a matrix equation and the response function is given by
\begin{equation}
	\chi_{\text{TDA}}^{}(q,\omega )=\chi_+(q,\omega )+\chi_-(q,\omega ), 
\end{equation}
with $\chi_{\pm}(q,\omega)$ as defined above, i.e. simply the HF response function if only 1-\emph{ph} contributions are included.

\begin{figure}[t]
	\centering 
	\includegraphics[width=0.95\linewidth, height=0.25\textheight]{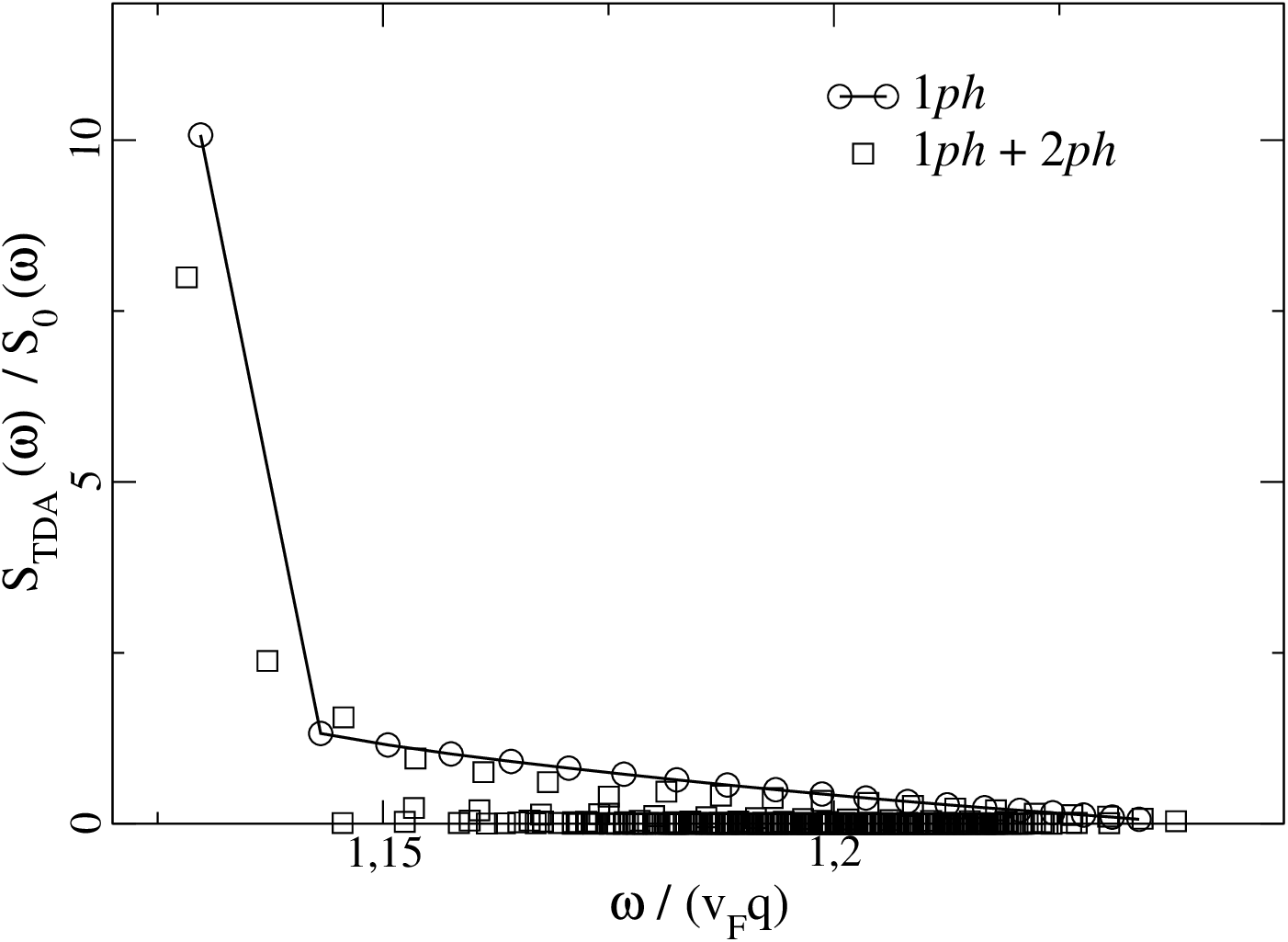}
	\caption{Response function for the spinless model in TDA including only 1-\emph{ph}(circles)- and 1-\emph{ph}+2-\emph{ph}(squares) contributions for $\a_4=0.2$ and quadratic interaction for constant momentum $q/(2k_F)=0.1$. The additional 2-\emph{ph}-terms result in slight deviations only. Calculations for $m_q=2\pi q/L=20$.}
	\label{fig:LR_2phTDA}
\end{figure}

For the 2-\emph{ph}--TDA the procedure is conceptionally the same, but more complicated matrix elements like $\bra{\a'\b'\m'\n'}(H-E_0\hf)\ket{\a\b\m\n}$ have to be calculated. Using 
\begin{equation*}\bra{\a'\b'\m'\n'}(H-E_0\hf)\ket{\a\b\m\n} = \bra{\a'\b'\m'\n'}[H,c_{\alpha}\dag c_{\b}\dag c_{\mu}c_{\n}]\ket{\text{HF}}\end{equation*} the commutator and the resulting matrix elements have to be calculated. Some details of this computation are presented in the appendix, we only give numerical results and some general remarks in this section. The charge and spin response function keep the form given in Eq.~(\ref{eq:ChiTDA}) but the functions $\chi_{\pm}$ differ, since the dimension of $a$ is much larger now. Figure~\ref{fig:LR_2phTDA} shows a comparison of the response function for the spinless model with 1-\emph{ph}  and 1- and 2-\emph{ph} excitations included for $\a_4=0.2$ and $\tilde q= 0.1$. 
 For small $\a_4$ the inclusion of 2-\emph{ph} contributions only slightly changes the resulting response function. Most of these additional excitations have very small weight and higher order excitations do contribute even less. 

\subsection{Discussion}
The $g_4$-model is quite remarkable considering the fact that the interacting ground state is still a HF state. Concentrating on low-lying excited states we can use the \mbox{1-\emph{ph}} TDA to calculate response functions and obtain numerical exact results for small momenta. The results for both charge and spin response function are analytic for an interaction constant in momentum space, and for finite interaction of arbitrary form only a numerically undemanding matrix diagonalisation is necessary. Since we are interested in changes due to finite curvature we focus our discussion on the DSF where these effects are more obvious than in the real part of the response functions.\\
The single-\emph{ph} excitations play the dominant role for weak interaction in a finite systems. Higher order excitations do of course exist, but their influence leads only to small corrections as can be seen in Figure~\ref{fig:LR_2phTDA}. Endorsed with this control over the quality of our approximations we focus for the following discussion on these most dominant contributions only.  

\begin{figure}[t]
	\centering 
 	\includegraphics[width=0.95\linewidth, height=0.25\textheight]{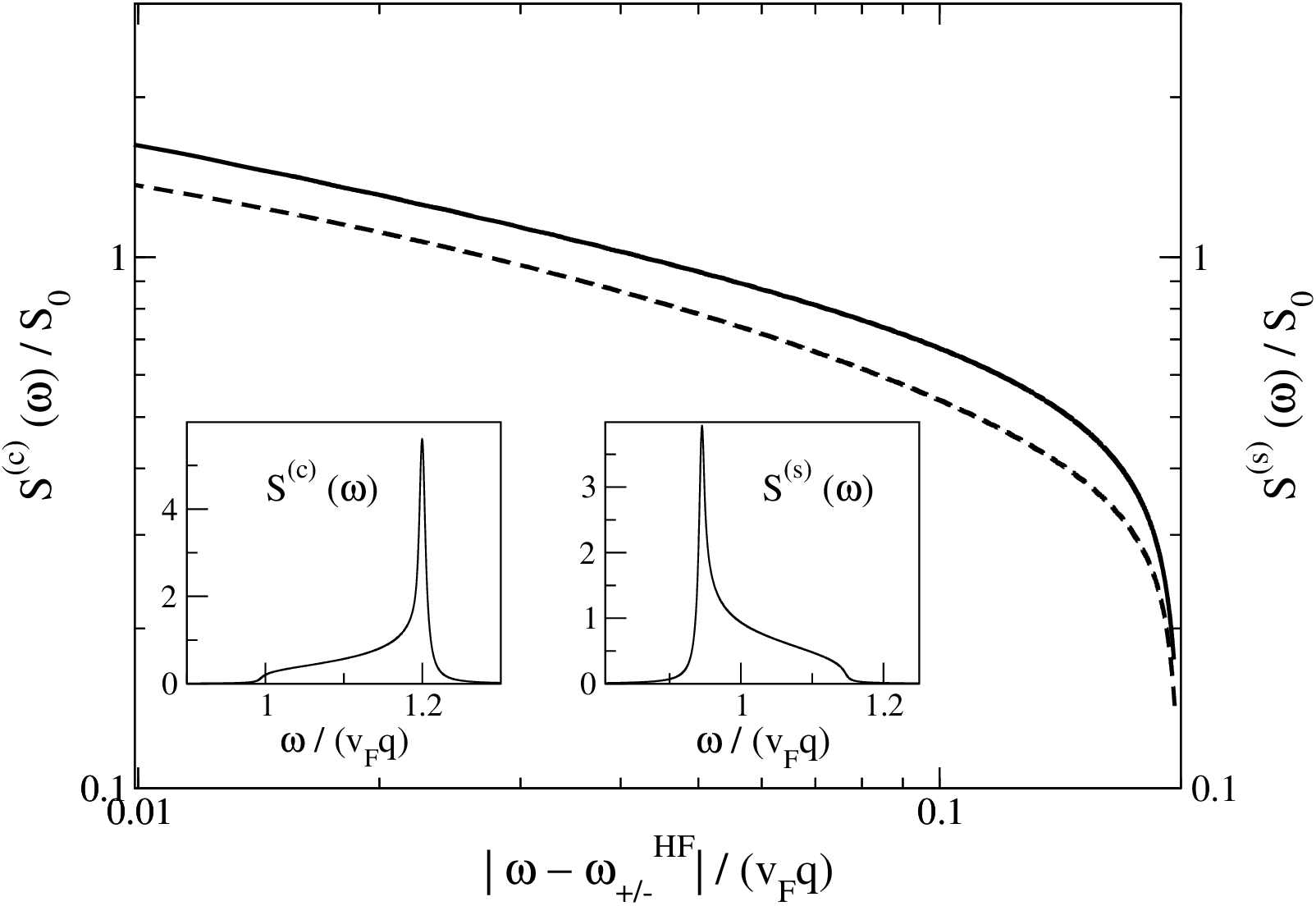}
	\caption{Charge (dashed) and spin (full line) response function in 1-\emph{ph} TDA close to the upper or lower boundary of the HF continuum in log-log plot for $\tilde q = 0.1$ and quadratic interaction with $\a_4=0.1$ (charge response) and $\a_4=0.05$ (spin response), respectively. A power-law behaviour close to the respective boundary is discernible. Insets: Charge and spin response function. Calculations for $m_q=2\pi q/L=20$ and $\eta=5\cdot10^{-3}$.}
	\label{fig:power-law_TDA}
\end{figure}

For fixed momentum $q$ one finds a transition from the box-like shape of the DSF in the non-interacting case to the dominance of a delta peak corresponding to the plasmon of the TL model. As already pointed out by Pirooznia and Kopietz\cite{Kopietz} there are two relevant limits: 1)~vanishing curvature $\tilde q=q/(2mv_F)\ll1$ or strong interaction $\a\gg1$ and 2)~strong curvature $\tilde q\gg1$ or vanishing interaction $\a\ll1$. For the shape of the DSF only the ratio $x\equiv\a/\tilde q$ of the two dimensionless quantities is important, leading to a box-like shape for $x\ll1$ and to a dominant plasmon peak for $x\gg1$.
In the parameter regime between these simple limits a quite peculiar shape appears for intermediate ratios $x\approx1$, the details depending on the type of interaction under consideration (for a comparison of different interaction forms for otherwise constant parameters \emph{cf.}~Figure~\ref{fig:LR_RPAE_WW} below). The DSF exhibits a divergence at the upper (lower) boundary of the HF continuum for the charge (spin) response function. This shows up as an emerging peak growing in weight compared to the continuum with increasing interaction strength. Figure~\ref{fig:power-law_TDA} shows a log-log plot of the charge and spin response function in 1-\emph{ph} TDA for $x=1$ and $x=1/2$, respectively. The two functions exhibit a power-law divergence at the respective boundary where the frequency is determined by the zero of the denominator in Eq.~(\ref{eq:ChiTDA}). This picture breaks down for $x\gg1$ and in particular for too strong interaction. The peak then separates from the HF continuum and power-law characteristics is clearly absent. Note that the necessary use of a finite imaginary part turns the delta function peaks into Lorentzians. These have a width of the order of $\eta$, therefore a log-log plot is only reasonable down to that scale, since the explicit form of the Lorentzians appears below it. 

\begin{figure}[t]
	\centering 
 	\includegraphics[width=0.95\linewidth, height=0.25\textheight]{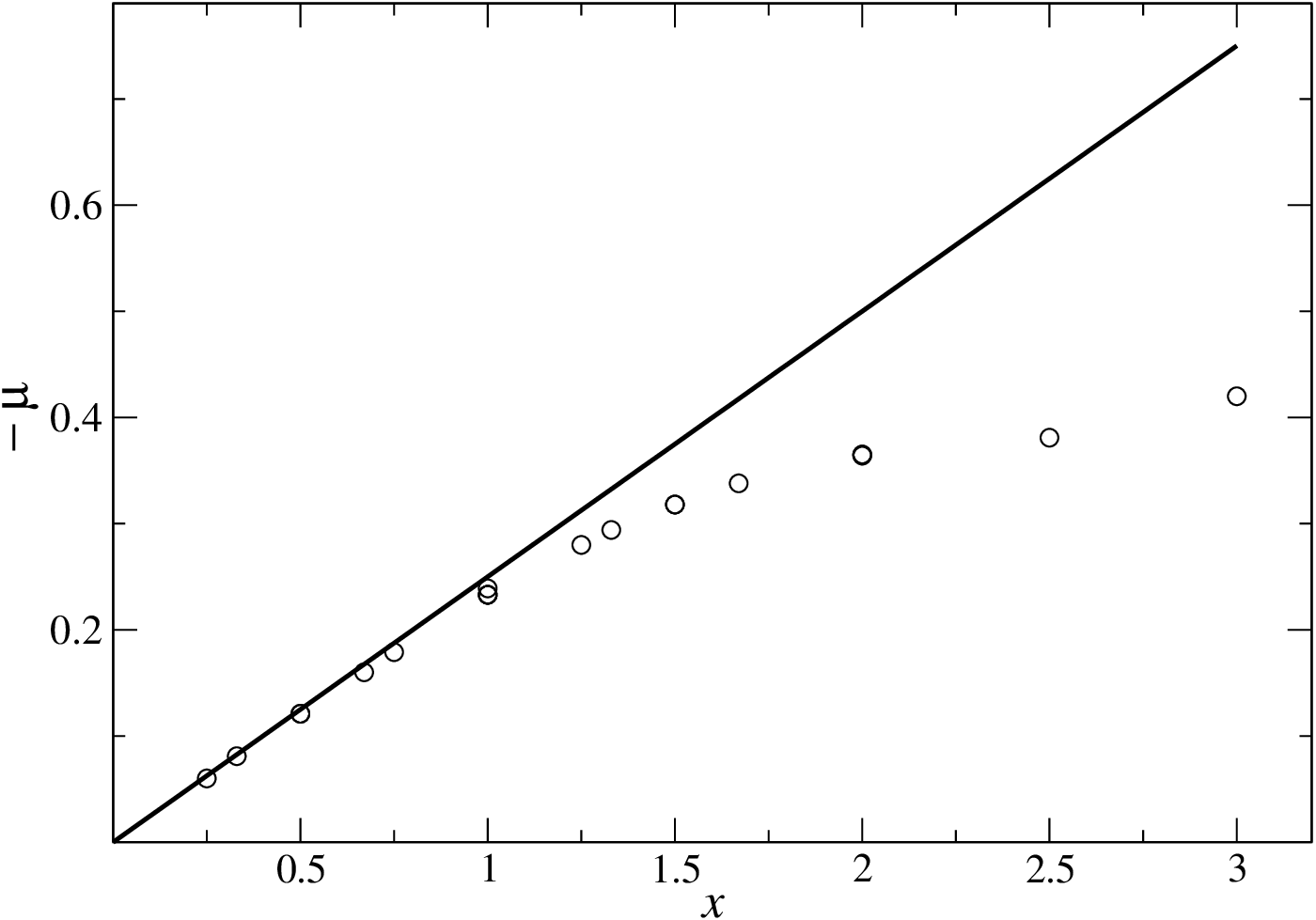}
	\caption{Exponent for the divergence of the DSF for the spinless model as predicted by Pustilnik \emph{et al.}\cite{Russen} (full line) according to Eq.~(\ref{eq:exponent}) and extracted from log-log plots within 1-\emph{ph} TDA (circles).}
	\label{fig:exponent}
\end{figure}

A similar behaviour holds in the spinless case. A transition from the non-interacting case to a plasmon peak of the TL model at a frequency approaching $\omega _{q}^{\text{TL}}$ with increasing interaction takes place. For intermediate ratios of $\tilde q$ and $\a_4$ a clear power-law behaviour shows up (\emph{cf.} Figure~\ref{fig:power_law} below). Pustilnik~\emph{et al.}\cite{Russen} predicted a power-law divergence for the spinless model close to the lower boundary $\omega _-$ of the non-interacting continuum for the DSF given by
\begin{equation}\label{eq:exponent}
	\frac{S(q,\omega )}{S_0(q,\omega )}=\left[\frac{2\tilde q}{\tilde\omega -\tilde\omega _{-}}\right]^{\mu}, \text{ for }0<\omega -\omega _-\ll q^2/m
\end{equation}
with an exponent $\mu = \frac{m_e}{\pi q}[g_4(0)-g_4(q)]\propto x = \a_4/\tilde q$ for a short-ranged quadratic interaction $g_4(q) = g_4(0)[1-q^2/k_c^2]$ and independent of the inter-band interaction $g_2$. The 'edge' of possible excitations is given by the lower bound of the HF continuum $\omega _-^{\text{HF}}$ in our approximation, and can be given explicitly for constant interaction $g_i=g_i(0)\Theta(q-|k_c|)$ where it reads $v_Fq(1+\a_4-\tilde q)$. In contrast to Pereira \emph{et al.}\cite{Spins} the exponents we find are not consistent with the above mentioned predictions but are usually smaller. This is depicted in Figure~\ref{fig:exponent} where the exponent extracted from log-log plots of the DSF in 1-\emph{ph} TDA is shown for different values of the ratio $x = \a_4/\tilde q$ along with the expected values according to Eq.~(\ref{eq:exponent}). For very small $x$, i.e. closer resemblance to the non-interacting case, the agreement is good, but clearly deteriorates for $x>1$. Note that this is not compensated by 2-\emph{ph} contributions since they only tend to further decrease the exponent. In this respect our results confirm the existence of a power-law, but with different exponents and only within a limited parameter regime where $x$ is of order unity.

\section{Inter- and intra-band interactions: the full model}\label{sec:full_model}
The previous section treated intra-band interactions only, i.e. we considered the pure $g_4$-model. We now drop this limitation and take also inter-band interactions $g_2$ into account. We treat this full model within linearised TDHF usually referred to as \emph{random phase approximation with exchange} (RPAE). This amounts to a clear improvement over single \emph{ph} TDA. Correlations in the ground state are now included and the excited states contain more correlations in a form that particle-hole states can not only be created but also annihilated\cite{RingSchuck} or as Fetter and Walecka put it: \emph{``TDA has one and only one particle-hole pair present at any instant of time, whereas the RPA permits any number of particle-hole pairs to be present simultaneously.''}\cite{FetterWalecka}
We first introduce the formalism, then derive our results for the charge and spin response function and close this section with a discussion of the latter.

\begin{widetext}
\subsection{Formalism}\label{sec:RPAE}
The change in particle-hole expectation values $\delta\exv{c_{k_a,\sigma}\dag c_{k_a+q,\sigma}^{}}$ in linear order due to an external time-dependent potential having Fourier components $V_{k_a+q,k_a}^{\sig}(\omega )=V_{a}^{\sig}(q,\omega )/L$ can be written\cite{KS}
	\begin{multline}
	\delta\exv{c_{k_a,\sigma}\dag c_{k_a+q,\sigma}^{}}= \frac{1}{L}\frac{f(\eps_{k_a}\hf)-f(\eps_{k_a+q}\hf)}{\omega-(\eps_{k_a+q}\hf-\eps_{k_a}\hf)}\Bigl\{V_{a}^{\sigma}(q,\omega)  +\sum_{k''\sigma''}g_2(q)\delta\exv{c_{k_{\bar a}'',\sig''}\dag c_{k_{\bar a}''+q,\sig''}^{}}\\
	+\sum_{k''\sigma''}[g_4(q)-\delta_{\sig\sig''}g_4(k_a-k_a'')] \delta\exv{c_{k_a'',\sig''}\dag c_{k_a''+q,\sig''}^{}} \Bigr\},
	\end{multline}
from which we get an equation for the response function matrix elements 
$\chi_{kk'}^{\sig\sig'}(q,\omega)\equiv \frac{\partial \delta\exv{c_{k,\sig}\dag c_{k+q,\sig}^{}}}{\partial V_{k'+q,k'}^{\sig'}}$.\cite{endnote}
In form of a matrix equation using the projection technique it reads 
	\begin{equation}\label{eq:ChiRPAE_Gross}
	\left(\begin{array}{cccc}\omega-a & -b_4& -b_2& -b_2\\-b_4 & \omega -a & -b_2& -b_2\\
				 -b_2& -b_2 & -\omega -a & -b_4\\ -b_2& -b_2& -b_4 & -\omega -a\end{array}\right)
	\left(\begin{array}{cccc}
		\chi^{++}_{PP} & \chi^{+-}_{PP} & \chi^{++}_{PQ} & \chi^{+-}_{PQ}\\ 
		\chi^{-+}_{PP} & \chi^{--}_{PP} & \chi^{-+}_{PQ} & \chi^{--}_{PQ}\\
		\chi^{++}_{QP} & \chi^{+-}_{QP} & \chi^{++}_{QQ} & \chi^{+-}_{QQ}\\
		\chi^{-+}_{QP} & \chi^{--}_{QP} & \chi^{-+}_{QQ} & \chi^{--}_{QQ}\\\end{array}\right)
	= \text{diag}(1).
	\end{equation}
\end{widetext}
Introducing matrices $(A)_{k,k'}^{\sig\sig'}=(\eps_{k+q}\hf-\eps_k\hf)\delta_{kk'}\delta_{\sig\sig'} -[g_4(q)-\delta_{\sig\sig'}g_4(k-k'')]$ and $(B_2)_{k,k'}^{\sig\sig'}=g_2(q)$ the last equation takes a simpler form 
	\begin{equation}\label{eq:ChiRPAE_Klein}
	\left(\begin{array}{cc}\omega-A & -B_2\\-B_2 & -\omega -A \end{array}\right)
	\left(\begin{array}{cc}\chi_{PP} & \chi_{PQ}\\\chi_{QP} & \chi_{QQ} \end{array}\right)
	= \left(\begin{array}{cc} 1 & 0 \\ 0 & 1 \end{array}\right).
	\end{equation}
To calculate the charge and spin response function we use the states $\ket{s}=\ket{+}+\ket{-}$ and $\ket{a}=\ket{+}-\ket{-}$ defined in Sec.~\ref{sec:RF}. We thus have
	\begin{equation}
	\chi^{(c)}_{ij}=\bra{s}\chi_{ij}\ket{s} \quad\text{and}\quad \chi^{(s)}_{ij}=\bra{a}\chi_{ij}\ket{a} 
	\end{equation}
where $i,j$ equal $P$ or $Q$. The block matrix $B_2$ can accordingly be written as $B_2=g_2(q)\ket s\bra s$. 

\subsection{Results}

\begin{figure}[th]
	\centering
	\includegraphics[width=0.95\linewidth, height=0.25\textheight]{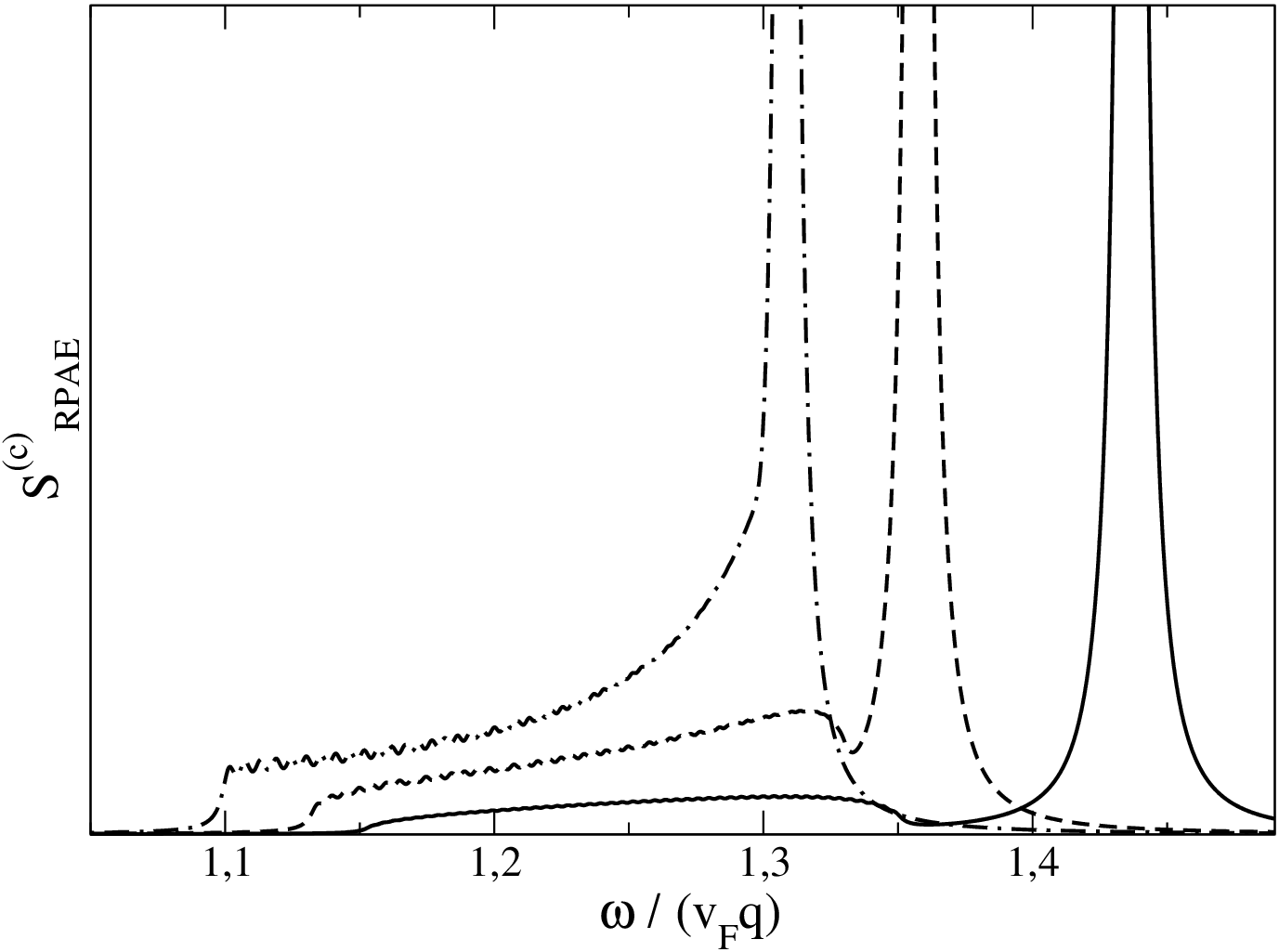}
	\caption{Charge response function in RPAE for constant (full line) $v_q=v_0\Theta(q-|k_c|)$, quadratic (dashed) $v_q=v_0(1-q^2/k_c^2)$ and exponential (dashed-dotted) interaction $v_q=v_0\exp(-|q/k_c|)$. Dimensionless interaction strength $\a=0.25$ and momentum $\tilde q= q/(2k_F)=0.1$ which corresponds to an intermediate relation between interaction and curvature of $x=2.5$. Calculations for $m_q=50$ and finite width of the peaks results from a finite imaginary part $\eta = 2.5\cdot10^{-3}$.}
	\label{fig:LR_RPAE_WW}
\end{figure}

\setcounter{paragraph}{0}
For the charge response function Eq.~(\ref{eq:ChiRPAE_Klein}) can be solved easily. We get two systems of two coupled equations reading e.g. \begin{equation*}
	(\omega-A)^{-1}=\left[1+ (\omega -A)^{-1}B_2(\omega+A)^{-1}B_2\right]\chi_{PP}.
	\end{equation*}
Multiplication from left with $\bra{s}$ and from right with $\ket{s}$ yields with the simple structure of $B_2$ the $PP$-part (and $QP$-part as well) of the RPAE charge response. Noting that  
	\begin{equation}
	\bra{s}(\omega \pm A)^{-1}\ket{s}=\frac{2\chi_{\pm}}{1-g_4(q)\chi_{\pm}}=\chi^{(c)}_{g_4,\pm}(q,\omega)
	\end{equation}
the RPAE charge response reads
	\begin{equation}\label{eq:ChiC_RPAE}
	\chi^{(c)}_{\text{RPAE}}(q,\omega )=\frac{\chi_{g_4,-}^{(c)}+\chi_{g_4,+}^{(c)}+2g_2(q)\chi_{g_4,+}^{(c)}\chi_{g_4,-}^{(c)}}{1-[g_2(q)]^2\chi_{g_4,+}^{(c)}\chi_{g_4,-}^{(c)}}.
	\end{equation}
These expressions are analytic for a constant interaction and in general only the functions $\chi_{\pm}(q,\omega )$ have to be determined numerically. Note that $g_2$- and $g_4$-interaction enter expression~(\ref{eq:ChiC_RPAE}) independently and can therefore be treated completely independently. Figure~\ref{fig:LR_RPAE_WW} shows the charge response function in RPAE for the galilei-invariant model where $g_2(q)=g_4(q)=v(q)$ for various types of interaction, specifically for constant $v_q=v_0\Theta(q-|k_c|)$, quadratic $v_q=v_0(1-q^2/k_c^2)$ and exponential interaction $v_q=v_0\exp(-|q/k_c|)$. Note that these different forms of the interaction in momentum space do not lead to qualitatively different results, but the generic shape and behaviour of the DSF is similar for all of them. For pure intra-band interaction, i.e. $g_2\equiv0$ we recover the result of Sec.~\ref{sec:g4}.
 \\
For spinless electrons the matrix equation~(\ref{eq:ChiRPAE_Gross}) is considerably simplified by dropping the spin indices and one finds for the RPAE response function\cite{Diplomarbeit}
	\begin{equation}\label{eq:Chi_RPAE}
	\chi^{}_{\text{RPAE}}(q,\omega )=\frac{\chi_{-}^{}+\chi_{+}^{}+2g_2(q)\chi_{+}^{}\chi_{-}^{}}{1-[g_2(q)]^2\chi_{+}^{}\chi_{-}^{}}.
	\end{equation}
where $\chi_{\pm}(q,\omega )=\mp\sum_{kk'}[(\omega \pm a)^{-1}]_{kk'}$.

For the spin response function the components of the RPAE spin response function are needed, i.e. the functions $\bra{a}\chi_{ij}\ket{a}$ where $i,j$ equal $P$ or $Q$. Due to the symmetry of $A$ the non-diagonal elements of $\chi_{\text{RPAE}}^{(s)}$ vanish as the following reasoning shows. 
Consider $\chi_{PQ}=(\omega mega -A)^{-1}B_2\chi_{QQ}$ and multiply by $\bra{a}$ from left and $\ket{a}$ from right (analogous to the charge response), then 
	\begin{equation*}
		\bra{a}\chi_{PQ}\ket{a}=-g_2(q)\bra{a}(\omega -A)^{-1}\ket{s}\bra{s}\chi_{QQ}\ket{a}
	\end{equation*}
holds. The coefficient on the rhs vanishes, since the two diagonal elements $\omega -a$ of $\omega - A$ are equal and $A$ is symmetric. The same holds for the inverse and the expectation value $\bra{a}\cdot \ket{s}$ is nothing but the difference between the diagonal and off-diagonal elements.\cite{endnote0}

The RPAE spin response is thus block diagonal and only 
	\begin{equation}
	\left(\begin{array}{cc}\omega-A & -B_2\\-B_2 & -\omega -A \end{array}\right)
	\left(\begin{array}{cc}\chi_{PP} & 0\\0 & \chi_{QQ} \end{array}\right)
	= \left(\begin{array}{cc} 1 & 0 \\ 0 & 1 \end{array}\right).
	\end{equation}
has to be considered, yielding the simple solution
	\begin{equation}
	\chi^{(s)}_{\text{RPAE}} (q,\omega) = \frac{2\chi_-}{1+g_4(q)\chi_-} +\frac{2\chi_+}{1+g_4(q)\chi_+},
	\end{equation}
which coincides with the 1-\emph{ph} TDA spin response function. This result is again analytic for a constant interaction (and coincides with earlier results\cite{KS} in this case) and easy to calculate numerically for other types of interaction. Note that the spin response function is completely independent of inter-band interactions, i.e. independent of $g_2$ and therefore has to be the same as in Sec.~\ref{sec:g4}. For increasing ratio of the interaction strength $\a$ over the dimensionless momentum $\tilde q$ a collective spin mode emerges which carries most of the spectral weight and moves to the frequency predicted by the TL model $\omega _{q,s}^{\text{TL}}=v_Fq$. 

\subsection{Discussion}
\begin{figure}[th]
	\includegraphics[width=0.95\linewidth, height=0.25\textheight]{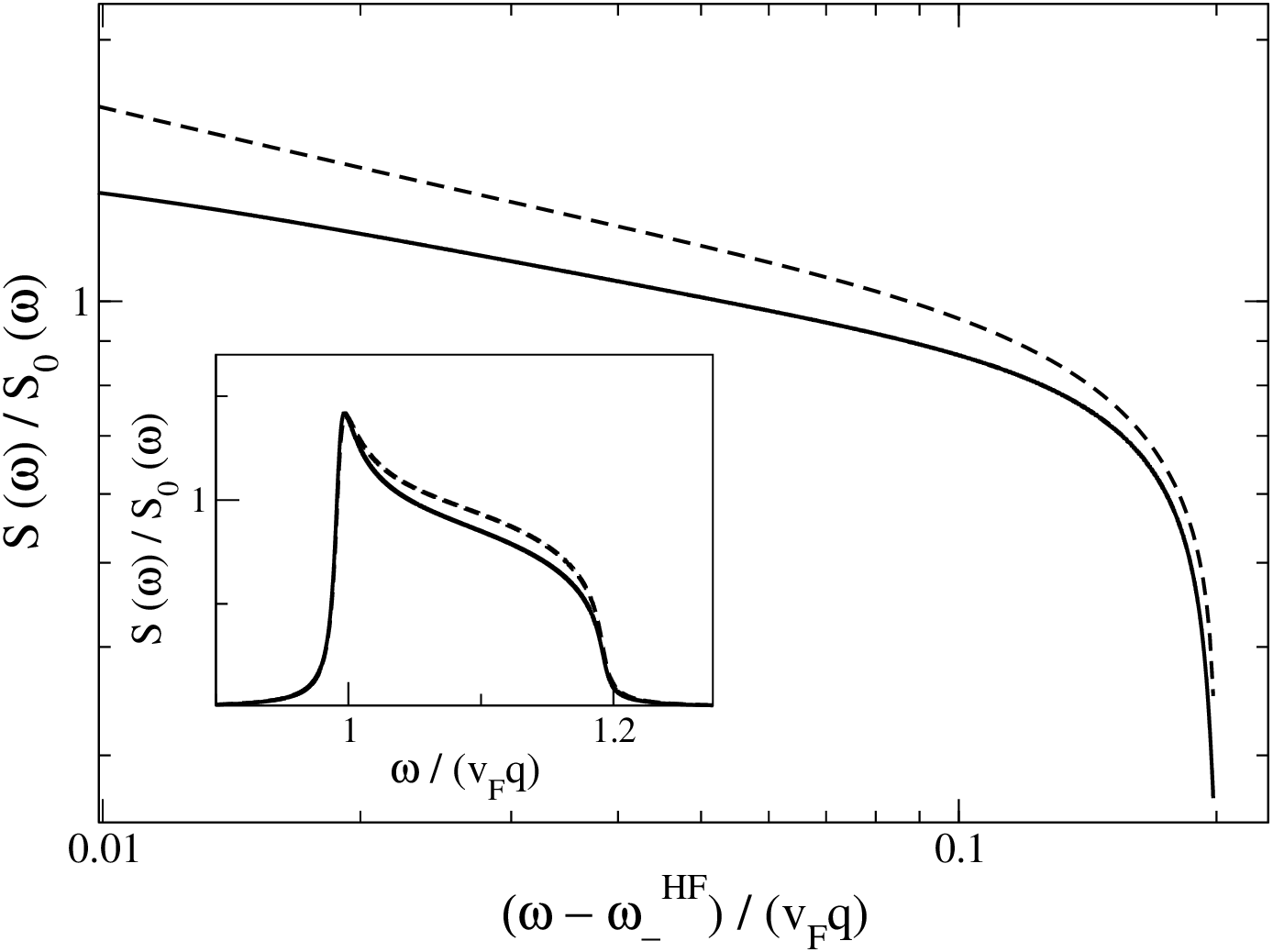}
	\caption{Dynamic structure factor for the spinless model in 1-\emph{ph}-TDA (dashed) for $\a_2=0.1, \a_4=0.0$ and RPAE (full line) for $\a_2=\a_4=0.1$, both for $q/(2k_F)=0.1$. Calculations for $m_q=100$ and $\eta=5\cdot 10^{-3}$.}
	\label{fig:power_law}
\end{figure}

The full model shows a similar behaviour as the $g_4$-model: In varying the ratio $x=\a/\tilde q$ of dimensionless momentum $\tilde q$ and interaction strength $\a_{2(4)}$ we find a transition from a box-like shape (for $x\gg1$ akin to the non-interacting case) and a dominant plasmon peak (for $x\ll1$ close to the predictions of the TL model). For growing interaction strength the peak approaches the frequency given by the TL model: $[\omega _{q,c}^{\text{TL}}]^2=(v_Fq)^2[(1+2\a_4)^2-(2\a_2)^2]$ for the charge mode and $[\omega _{q,s}^{\text{TL}}]^2=(v_Fq)^2$ for the spin mode.
Again we find strong indications of power-law behaviour for the charge and spin response function for intermediate ratios $x\approx1$, now not only depending on the form of the interaction under consideration but also on the strength of the inter-band interaction $g_2(0)$. This holds for the spinless model, too. Figure~\ref{fig:power_law} shows the DSF for the spinless $g_4$-model and the full spinless model for constant momentum $\tilde q= 0.1$. 
The inclusion of $g_4$-couplings changes the DSF quantitatively but not the overall shape. But in contrast to Pustilnik \emph{et al.}\cite{Russen} we find a clear dependence of the exponent on inter-band interaction $g_2$ in the spinless model. Furthermore their exponent $\mu=\frac{m_e}{\pi q}[g_4(0)-g_4(q)]$ vanishes for constant interaction, but the DSF in RPAE shows a qualitatively similar behaviour as for quadratic interaction including a power-law divergence. As already discussed for the $g_4$-model this is only valid for parameters with $x\approx1$ and the divergence goes over to the dominant plasmon peak for values of $x$ smaller than one.  

\section{Summary}\label{sec:summary}
We have discussed charge and spin response functions for interacting electrons in one spatial dimension. We went beyond the assumption of the TL model and included deviations from a strictly linear energy dispersion. We first focused on pure intra-band interactions ($g_4$-processes), where the ground state is known to be a simple Slater determinant and expanded the excited state in \emph{ph} excitations. Expressions for the response functions were derived and the DSF has been discussed. A clear transition from the non-interacting case to the TL model was found depending on the ratio of the dimensionless momentum and the interaction strength. For intermediate values indications of power-law divergences were found but with exponents deviating from existing predictions.\\
The inclusion of inter-band interaction in the full model was realised within the RPAE and leads to quantitative but no qualitative changes. Again our examination of power-law exponents contradicted existing suggestions in the literature. 

\section*{Acknowledgements} The author is grateful to Sina Riecke, K$\bar{\text{a}}$rlis Mi\c{k}elsons, Christoph Karrasch, and Jens Birkholz for numerous discussions as well as to Kurt Sch\"onhammer for careful reading of the manuscript and many valuable hints throughout this work. The author acknowledges support within the framework of the Excellence Initiative by the German Research Foundation (DFG) through the Heidelberg Graduate School of
Fundamental Physics (grant number GSC 129/1). 

\appendix
\begin{widetext}
\section*{Appendix: TDA matrix elements}
The matrix elements for the 2-\emph{ph} TDA are
\begin{equation}\bra{\a'\b'\m'\n'}(H-E_0\hf)\ket{\a\b\m\n} = \bra{\text{HF}}	c_{\n'}\dag c_{\mu'}\dag c_{\b'}^{}c_{\alpha'}^{}[H,c_{\alpha}\dag c_{\b}\dag c_{\mu}c_{\n}]\ket{\text{HF}}.\end{equation}
The commutators are easily calculated: $[T,c_{\alpha}\dag c_{\b}\dag c_{\mu}c_{\n}] = (\eps_{\a}\hf + \eps_{\b}\hf - \eps_{\m}\hf - \eps_{\n}\hf)c_{\alpha}\dag c_{\b}\dag c_{\mu}c_{\n}^{}$ and 
\begin{equation}\begin{split}
	[V,c_{\alpha}\dag c_{\b}\dag c_{\mu}c_{\n}] & = \frac{1}{2}\Big\{\sum_{mnm'}\bar v_{mn,m'\a}c_{m}\dag c_{n}\dag c_{m'}c_{\m}
	 -\sum_{mm'n'}\bar v_{m\m,m'n'}c_{\a}\dag c_{m}\dag c_{n'}c_{m'}\Big\}c_{\b}\dag c_{\n}^{} \\
	& + \frac{1}{2}c_{\a}\dag c_{\m}^{} \Big\{\sum_{mnm'}\bar v_{mn,m'\b}c_{m}\dag c_{n}\dag c_{m'}c_{\n} -
	\sum_{mm'n'}\bar v_{m\n,m'n'}c_{\b}\dag c_{m}\dag c_{n'}c_{m'}\Big\}\end{split},
\end{equation}
We choose $\a>\b$ and $\m>\n$ (and similarly for other momenta) and find after some lengthy calculation the matrix elements
\begin{equation*}\begin{split}
	\bra{\a'\b'\m'\n'}(H-E_0\hf)\ket{\a\b\m\n} =  
	(\eps_{\a}\hf + \eps_{\b}\hf - \eps_{\m}\hf - \eps_{\n}\hf )\delta_{\a\a'}\delta_{\b\b'}\delta_{\m\m'}\delta_{\n\n'} 
	+ \bar v_{\a'\b',\a\b}\delta_{\m\m'}\delta_{\n\n'} + \bar v_{\m'\n',\m\n}\delta_{\a\a'}\delta_{\b\b'} \\
	-\delta_{\a\a'}\big[\bar v_{\b'\m,\b\m'}\delta_{\n\n'} - \bar v_{\b'\m,\b\n'}\delta_{\n\m'} +\bar v_{\b'\n,\b\n'}\delta_{\m\m'} -\bar v_{\b'\n,\b\m'}\delta_{\m\n'}\big]\\
	+\delta_{\a\b'}\big[\bar v_{\a'\m,\b\m'}\delta_{\n\n'} - \bar v_{\a'\m,\b\n'}\delta_{\n\m'} +\bar v_{\a'\n,\b\n'}\delta_{\m\m'} -\bar v_{\a'\n,\b\m'}\delta_{\m\n'}\big]\\
	-\delta_{\b\b'}\big[\bar v_{\a'\m,\a\m'}\delta_{\n\n'} - \bar v_{\a'\m,\a\n'}\delta_{\n\m'} +\bar v_{\a'\n,\a\n'}\delta_{\m\m'} -\bar v_{\a'\n,\a\m'}\delta_{\m\n'}\big]\\
	+\delta_{\b\a'}\big[\bar v_{\b'\m,\a\m'}\delta_{\n\n'} - \bar v_{\b'\m,\a\n'}\delta_{\n\m'} +\bar v_{\b'\n,\a\n'}\delta_{\m\m'} -\bar v_{\b'\n,\a\m'}\delta_{\m\n'}\big].	\end{split}
\end{equation*}
Moreover some mixed terms of the form $\bra{1ph}(H-E_0\hf)\ket{2ph}$ and $\bra{2ph}(H-E_0\hf)\ket{1ph}$ have to be calculated. Since the commutators are known already this can be done easily and one obtains
	\begin{gather*}
	\bra{\a'\b'\m'\n'}(H-E_0\hf)\ket{\a\m}  =  \bar v_{\m\b',\m'\n'}\delta_{\a\a'} -\bar v_{\m\a',\m'\n'}\delta_{\a\b'} +  \bar v_{\b'\a',\a\n'}\delta_{\m\m'} -\bar v_{\b'\a',\a\m'}\delta_{\m\n'}\\
	\bra{\a'\m'}(H-E_0\hf)\ket{\a\b\m\n}  =  \bar v_{\m'\b,\m\n}\delta_{\a\a'} - \bar v_{\m'\a,\m\n}\delta_{\a'\b} + \bar v_{\b\a,\a'\n}\delta_{\m\m'} -\bar v_{\b\a,\a'\m}\delta_{\n\m'}.
	\end{gather*}
With these matrix elements the response functions can be calculated numerically including 1-\emph{ph}- and 2-\emph{ph}-contributions.
\end{widetext}


\begin{thebibliography}{*}
\bibitem{endnote0}This can be seen explicitly for the solution of Eq.~(\ref{eq:Matrix_g4}), where $\chi^{++}=\chi^{--}$ and $\chi^{+-}=\chi^{-+}$ holds. Therefore \mbox{$\bra{a}(\omega -A)^{-1}\ket{s}=\chi^{++}-\chi^{--}+\chi^{+-}-\chi^{-+}=0$.}
\end{thebibliography}
\end{document}